\documentclass[reprint,aps,prl,superscriptaddress,showpacs,preprintnumbers]{revtex4-1}

\usepackage[usenames,dvipsnames]{color}
\usepackage{graphicx}
\usepackage{amsmath,amssymb,amsfonts}
\usepackage{braket}

\begin{document}

\title{Disorder-induced Localization in a Strongly Correlated Atomic Hubbard Gas}

\author{S.S. Kondov}
\altaffiliation[Now at: ]{Physics Department, Princeton University, Jadwin Hall, Princeton NJ 08544, USA.}
\author{W.R. McGehee}
\author{W. Xu}
\author{B. DeMarco}
\affiliation{Department of Physics, University of Illinois at Urbana-Champaign, Urbana, Illinois 61801, USA}
\date{\today}

\begin{abstract}
We observe the emergence of a disorder-induced insulating state in a strongly interacting atomic Fermi gas trapped in an optical lattice.  This closed quantum system free of a thermal reservoir realizes the disordered Fermi-Hubbard model, which is a minimal model for strongly correlated electronic solids.  In measurements of disorder-induced localization obtained via mass transport, we detect interaction-driven delocalization and localization that persists as the temperature of the gas is raised.  These behaviors are consistent with many-body localization, which is a novel paradigm for understanding localization in interacting quantum systems at non-zero temperature.
\end{abstract}
\maketitle

The impact of inter-particle interactions on localization of disordered quantum systems has been the subject of intense scrutiny for decades (see \cite{Finkelstein,Lee1985,PhysRevLett.81.4212,Miranda2005,Sanchez-Palencia2010} and references therein). Obtaining new insights into the interplay of interactions and disorder is critical to improving our understanding of quantum electronic solids such as the high-temperature superconducting cuprates and materials that exhibit colossal magnetoresistance, such as the manganites \cite{Miranda2005,Dagotto2005,Dagotto2005a}. Despite the application of a wide variety of sophisticated theoretical and numerical approaches, consensus regarding the nature of metal--insulator transitions and localization in strongly correlated systems has not been achieved.  A recent theoretical approach to these questions is many-body localization (MBL) \cite{Basko2006,PhysRevB.75.155111,PhysRevB.82.174411,nayak2013}, which overturns the conventional view holding that materials above zero temperature have non-zero conductivity in the presence of interactions. In a many-body localized state, a quantum system can remain an Anderson-localized insulator at non-zero temperature because the inter-particle interactions fail to generate thermally activated conductivity.


We investigate localization using an ultracold atomic gas trapped in a disordered optical lattice.  This precisely controllable system, which realizes the disordered Fermi-Hubbard model (DFHM) \cite{Belitz1994}---the minimal model for strongly correlated, disordered electronic solids---is free of a heat bath, such as phonons, that can lead to finite conductivity at nonzero temperature and foils direct tests of theories such as MBL in the solid state. The seminal theoretical work by Basko et al. on MBL \cite{Basko2006} explored the weakly interacting regime of a spinless DFHM; we investigate the strongly correlated limit which is challenging for theory and numerical approaches.  We probe disorder-induced metal--insulator transitions using mass transport measurements.  The disorder $\Delta_c$ required to localize the gas and produce an insulating state is determined for different ratios of the Hubbard interaction to tunneling energies.  We find that increased interactions stabilize the metal against localization and lead to an insulator--metal transition. We also show that localization occurs across a range of thermal energy densities at fixed disorder strength by varying the temperature of the gas.


In our experiment, fermionic $^{40}$K atoms cooled below the Fermi temperature $T_F$ and trapped in a cubic optical lattice potential formed from three pairs of counter-propagating laser beams play the role of the electrons in a solid \cite{EsslingerReview2010}.  The atoms are confined by a crossed-beam dipole trap that forms a parabolic potential as shown in Fig. 1.  An approximately equal mixture of two atomic hyperfine states ($\downarrow$=$|F=9/2,\allowbreak m_F=9/2\rangle$ and $\uparrow=\ket{F=9/2,m_F=7/2}$) are used to mimic the spin of the electron.

\begin{figure}[h!]
\includegraphics[width=0.45\textwidth]{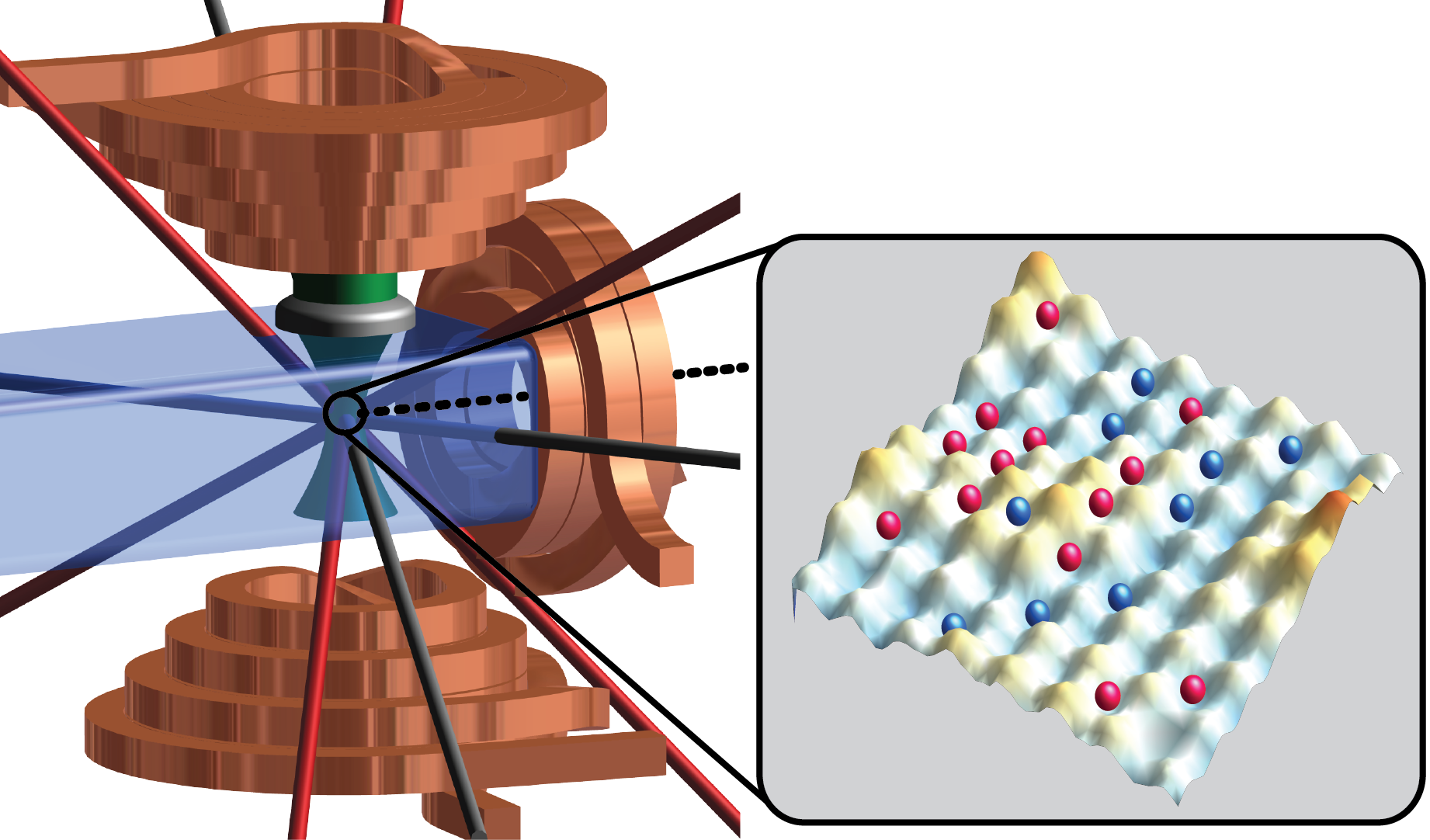}
\caption{Experimental geometry.  The atoms are cooled in a magnetic trap (copper) and an optical dipole trap formed from 1064~nm laser beams (gray lines).  Optical lattice laser beams (red) superimposed on the trap form a cubic lattice potential.  A 532~nm optical speckle field (green) is focused onto the atoms using a 1.1 $f$-number lens (gray hemisphere).  Atoms in two hyperfine states (red and blue spheres) are trapped in the disordered lattice potential (false color) formed at the intersection of all the laser beams.  The imaging direction is along the [111] direction of the lattice and is indicated by a dashed line.}
\label{fig:Fig1}
\end{figure}

In the lattice, the atoms tunnel between adjacent sites and two atoms on the same site (in different hyperfine states) interact through a low-energy \textit{s}-wave collision, thereby realizing the Fermi-Hubbard model (FHM) \cite{Jaksch1998,EsslingerReview2010}. Previous work with ultracold atoms has explored the Mott insulator (MI) phase \cite{Jordens2008,Schneider2008} and transport properties \cite{Strohmaier2007,Schneider2012} for the FHM. The equivalent of material parameters, such as the ratio $U/t$ of Hubbard interaction to tunneling energy, are precisely known and tunable over orders of magnitude by adjusting the power of the $\lambda=782.2$~nm lattice laser, which controls the lattice potential depth $s$.  We access the metallic phase in the lattice by employing a range of $s$ such that $U<12t$ and by adjusting the number of atoms $N$ and the geometric mean of the harmonic trap frequency $\omega$ so that the characteristic density $\tilde{\rho}=N\left(m\omega^2 d^2/12t\right)^{3/2}<5$ \cite{DeLeo2008}.  The trap leads to a spatially inhomogeneous density profile, with approximately 0.3--0.7 particles per site in the center of the clean lattice for each spin state \cite{SOM}.

By disordering the lattice potential using op\-ti\-cal speck\-le \cite{White2009,Pasienski2010}, we explore the DFHM with ultracold atoms for the first time.  The optical speckle field is produced by passing a 532~nm laser beam through a holographic diffuser and focusing it onto the atoms, as in Refs.~\citenum{White2009}, \citenum{Pasienski2010}, and \citenum{Kondov2011}. The atoms experience a potential proportional to the optical speckle intensity, which varies randomly in space. The strength of this disorder is characterized by the average disorder potential energy $\Delta$ and can be adjusted by varying the 532~nm laser power.  In contrast with experiments on solids, the disorder is precisely known (via optical microscopy) and continuously tunable, from complete absence to the largest energy scale present.

The disorder causes the ``clean" Hubbard model occupation $\epsilon$, interaction $U$, and tunneling $t$ energies to vary from site to site in the lattice. The atoms therefore realize a single-band DFHM described by the Hamiltonian $H=\sum_i U_i\hat{n}_{i\uparrow}\hat{n}_{i\downarrow}-\sum_{\left\langle ij\right \rangle,\sigma}t_{ij}\left(\hat{c}_{j\sigma}^{\dagger}\hat{c}_{i\sigma}+h.c.\right)+\sum_{i,\sigma}\left(\epsilon_i+m\omega^2 r_i^2/2\right)\hat{n}_{i,\sigma}$, where $i$ indexes the lattice sites, $\hat{c}_{i\sigma}^{\dagger}$ is the operator that creates an atom on site $i$ in spin state $\sigma=\uparrow,\downarrow$, $\langle ij\rangle$ indicates a sum over adjacent sites, $m$ is the atomic mass, $r_i$ is the distance from the trap center to site $i$, and $\hat{n}_{i,\sigma}=\hat{c}_{i\sigma}^{\dagger}\hat{c}_{i\sigma}$ is the number operator.  We work at sufficiently low temperature such that the atoms occupy only the lowest energy band.  The statistical distributions of Hubbard parameters are given in Refs.~\citenum{White2009} and \citenum{PhysRevA.81.013402}; the standard deviation of the $\epsilon_i$ distribution is approximately equal to $\Delta$.  We cite the Hubbard energies and $\Delta$ in units of the atomic recoil energy $E_R=h^2/8md^2\approx k_B\cdot 390 \text{nK}$, where $d=\lambda/2$ is the lattice spacing, and $h$ and $k_B$ are Planck's and Boltzmann's constants.

To study the influence of interactions and disorder on transport, we measure the response of the atomic quasimomentum distribution $n(q)$ to an applied impulse. We developed this method to measure disorder-induced localization for the Bose-Hubbard model in previous experiments \cite{Pasienski2010} and achieved quantitative agreement with quantum Monte-Carlo simulations \cite{Gurarie2009}.  An external force is applied to the gas by turning on a magnetic field gradient for 2~ms, which is short compared with the confining trap period \cite{SOM}.  Immediately following the impulse, the lattice is turned off in 200~$\mu$s, and we measure $n(q)$ by bandmapping and absorption imaging after 10~ms time-of-flight \cite{McKay2009}.  The center-of-mass (COM) velocity $v_{COM}$ of $n(q)$ is determined by measuring the displacement of the centroid of the imaged density profile from the case without an impulse.

In the metallic phase, applying an external force induces a COM velocity, which is manifest as an asymmetry in $n(q)$ and $v_{COM}\neq0$ (Fig. 2(a),i).  We observe that the introduction of disorder obstructs transport, leading to a localized insulating phase (Fig. 2(a),ii).  Low-temperature (i.e., $0.16\pm0.01$~$T_F$ in the trap) data taken for a range $s=4$--7~$E_R$ (corresponding to $U/12t\approx0.20$--0.75) and $\Delta\approx$0--1.5~$E_R$ are shown in Fig. 2(a).  At all lattice potential depths, increasing $\Delta$ causes $v_{COM}$ to decrease.

\begin{figure}[h!]
\includegraphics[width=0.45\textwidth]{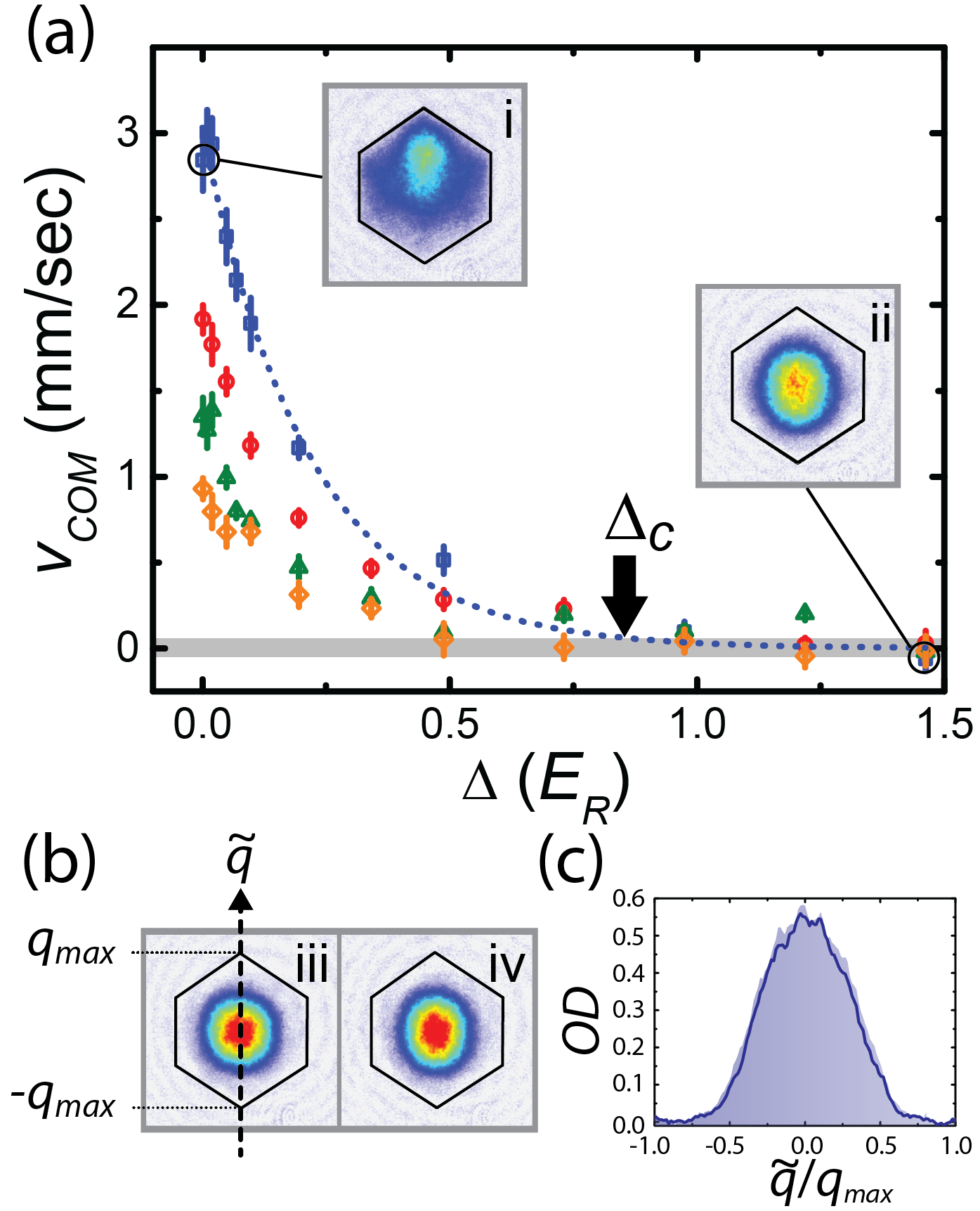}
\caption{(a) The COM velocity of the atom gas measured after an applied impulse for $s=4$ (blue squares), 5 (red circles), 6 (green triangles), and 7~$E_R$ (orange diamonds). Sample images used to determine $v_{COM}$ are shown in false color for $s=4$~$E_R$ for $\Delta=0$~$E_R$ (i) and $\Delta=1.46$~$E_R$ (ii). The field-of-view for all absorption images used in this work is 0.54~mm.  The projection of the Brillouin zone onto the imaging plane, which is a hexagon because of the imaging and lattice-beam geometry, is indicated using solid black lines.  The blue dotted line is the exponential fit used to determine $\Delta_c$ for $s=4$~$E_R$; the arrow indicates $\Delta_c$ for $s=4$~$E_R$.  The error bars are the standard error in the mean for the 7--9 experimental runs that are averaged for each data point. (b) Images taken at $s=4$~$E_R$, $\Delta=0$~$E_R$ (iii) and $\Delta=1.46$~$E_R$ (iv) without an impulse.  The quasimomentum $\tilde{q}$ projected along the vertical axis in the imaging plane is measured in units of the maximum allowed quasimomentum in the Brilloun zone $q_{max}$.  (c) Traces through images (iii) (solid blue line) and (iv) (blue shaded region) showing the measured optical depth (OD).}
\label{fig:Fig2}
\end{figure}


Sufficient $\Delta$ to completely arrest motion, signifying a metal--insulator transition, is achieved for all $s$. Localization in three dimensions has been previously observed for non-lattice gases using weakly \cite{Jendrzejewski2012} and non-interacting \cite{Kondov2011} atoms.  The metal smoothly transforms to the insulating state because the entire continuous energy spectrum of non-localized, single-particle states in the gas can contribute to $v_{COM}$.  As revealed in the images and profiles shown in Figs. 2(b) and 2(c), disorder has a minor impact on $n(q)$.  The localized insulating state we observe emerge is therefore qualitatively distinct from a band or Mott insulator---the quasimomentum distribution is narrow and states proximate in quasimomentum to the band edge are unfilled.  Based on the rms size of the image shown in Fig.~2(b) and assuming an exponential density distribution for localized states, we estimate a lower limit of 2.5 sites for the average localization length.


To quantitatively identify the transition to a localized state, we measure the characteristic disorder $\Delta_c$ required to completely arrest motion.  In a single-band system, all energy scales are bounded and a finite disorder strength localizes all (single-particle) states.  The $\Delta_c$ we measure corresponds to the average disorder potential energy required for the mobility edge trajectory to traverse the band and localize all states in the non-interacting limit.  In the Anderson model, $\Delta_c$ is a fixed fraction of the total bandwidth, and is $6t$ ($16t$) for Gaussian (uniformly) distributed site energies \cite{Bulka}.  A calculation of the mobility edge trajectory for the lattice we employ, which has a distribution of site energies sharing properties of both the Gaussian and uniform cases, is unavailable.

We measure $\Delta_c$ as the disorder potential energy necessary to eliminate $v_{COM}$ within the experimental resolution $v_{res}=0.05$~mm/s.  The grey band in Fig. 2(a) shows $v_{res}$, which is the standard error of the mean in $v_{COM}$ when an impulse is not applied.  $\Delta_c$ is determined from a heuristic fit used to smooth the data at fixed $s$ to $v_{COM}=A e^{-\Delta/\Delta_c log\left(A/v_{res}\right)}$ with $A$ and $\Delta_c$ as free parameters.  The resolution $v_{res}$ corresponds to a 10~pK thermal velocity, which is three orders of magnitude smaller than all other energy scales---the Fermi temperature is approximately 200~nK, the temperature of the gas is roughly 30~nK, and $\Delta_c/k_B\sim200$--300~nK.  Our determination of $\Delta_c$ is thus an excellent approximation to the disorder required to localize the band.


By comparing $\Delta_c$ measured at different $s$ and interaction strengths, we observe an interaction-driven insulator--metal (i.e., delocalization) transition. Fig. 3 shows $\Delta_c$ normalized by $12t$ for the $U/12t$ corresponding to each lattice potential depth sampled in Fig. 2.  The characteristic disorder and Hubbard energies are normalized to $12t$, which is the maximum kinetic energy for single particles.  We observe that as $U/12t$ increases, $\Delta_c/12t$ increases, and thus a localized insulating phase can be transformed to a metallic state by stronger interactions.  For a non-interacting system, $\Delta_c/12t$ would remain fixed as $s$ was varied because the bandwidth $12t$ (controlled solely by $s$) determines all energy scales \cite{Bulka}. The slope of a linear fit to the data is positive at greater than the six-standard-deviation level.


To exclude the observed metal--insulator transition as a percolation transition, in which particles are classically confined to finite spatial region of energetically accessible sites, we determine the percolation threshold \cite{Gurarie2009}, which is shown in Fig. 3 as a dashed line and exceeds the measured $\Delta_c/12t$ by an order of magnitude at high $U/12t$.  The $\Delta_c/12t$ we measure is approximately a factor of 3--4 smaller compared with a statistical dynamical mean-field theory (SDMFT) prediction for a metal-insulator transition \cite{Semmler2010}.  This discrepancy may be explained by the Bethe lattice geometry used in the SDMFT, uncontrolled approximations employed in SDMFT, and the presence of the trap in the experiment.

\begin{figure}
\includegraphics[width=0.45\textwidth]{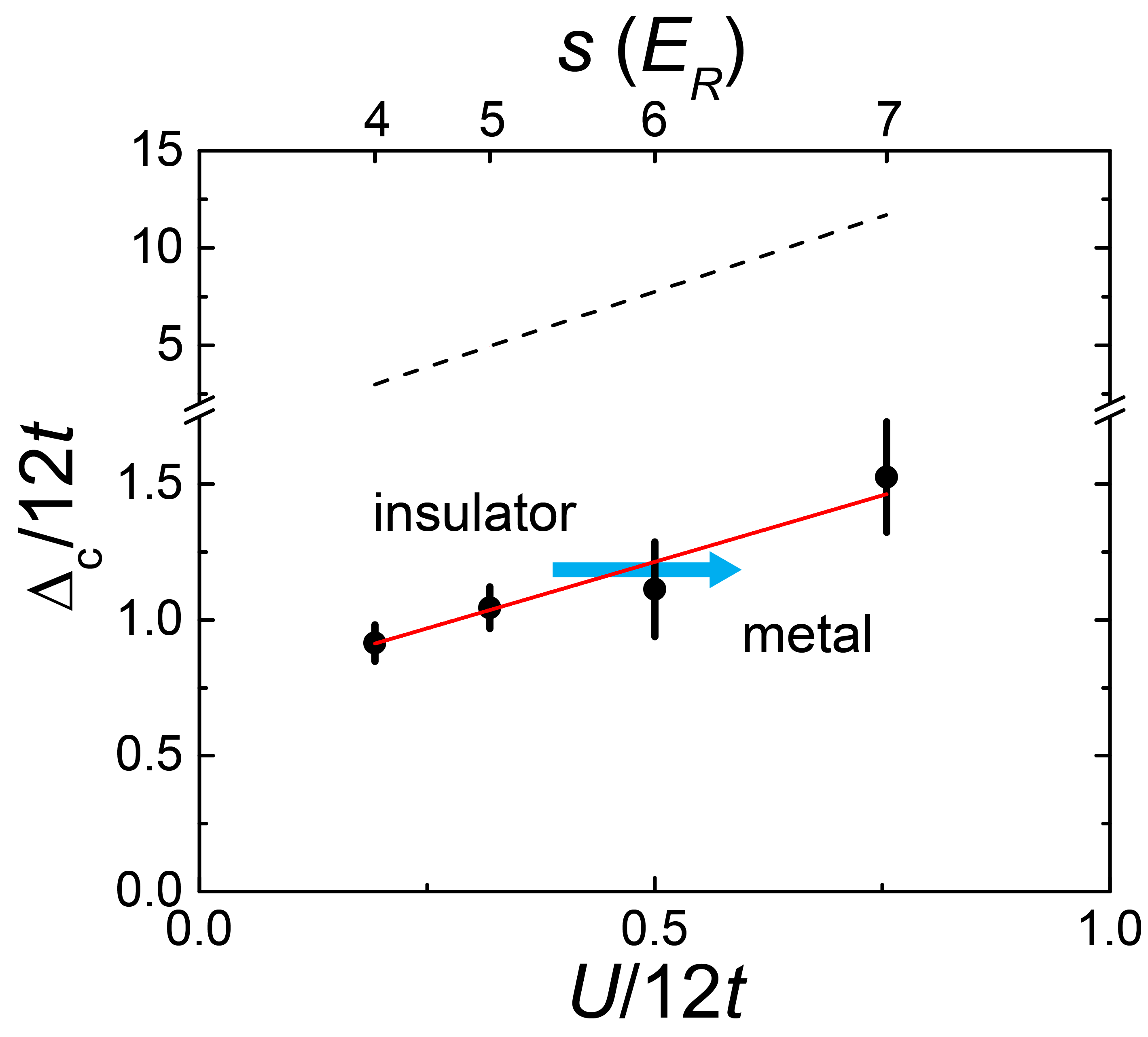}
\caption{Interaction induced delocalization. The characteristic disorder strength $\Delta_c/12t$ is shown for varying interaction strength $U/12t$, which is controlled by tuning the lattice potential depth $s$. The blue arrow indicates an interaction-driven insulator--metal transition.  $\Delta_c$ varies by less than 20\% across this range, while $t$ changes by a factor of 2.2. The error bars show the uncertainty in the fit to the data in Fig. 2 used to determine $\Delta_c$.  The percolation threshold is shown as a dashed line, and a linear fit to the data as a red solid line.}
\label{fig:Fig3}
\end{figure}

To explore the temperature dependence of localization, we vary the temperature of the gas before turning on the disordered lattice with $s=4$~$E_R$ and fixed $\Delta$.  In Fig.~4, we show measurements of $v_{COM}$ for temperatures ranging from 40 to 150~nK (corresponding to an entropy of $k_B\times\left(1.9\text{--}3.3\right)$ per particle) in the harmonic trap.  In order to account for the compression of this temperature range in the lattice because of the maximum kinetic energy attached to the finite bandwidth, we treat the disorder as an overall chemical potential shift and estimate the corresponding temperature in the lattice $T_{lat}$ using a self-consistent Hartree-Fock calculation to match entropy \cite{SOM}.  We fix $\Delta=1$~$E_R$, which is the characteristic disorder $\Delta_c$ for localization at the lowest temperature, in order to maximize the sensitivity to temperature.  For reference, we also show data with motion present for $\Delta=0.4$~$E_R$.

\begin{figure}
\includegraphics[width=0.45\textwidth]{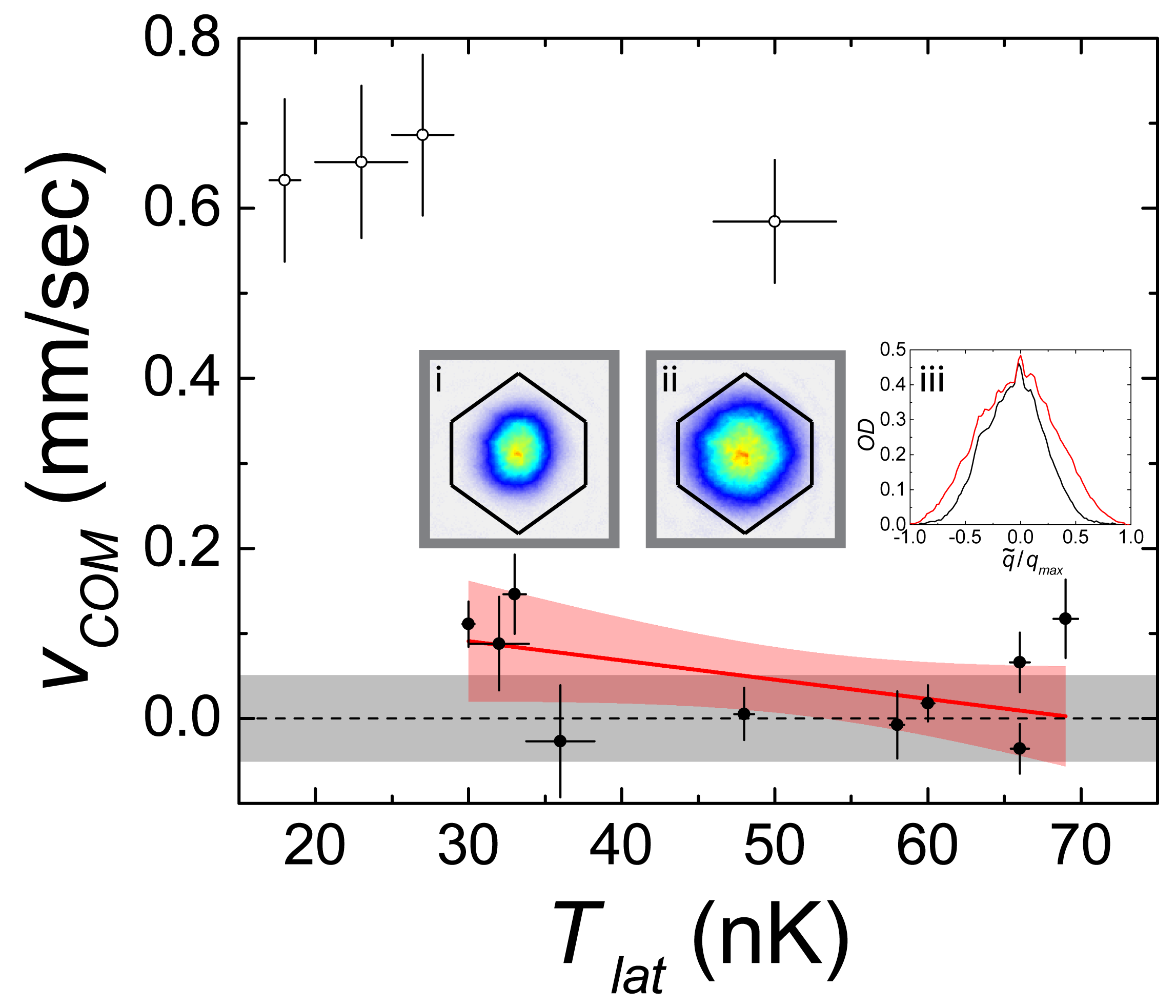}
\caption{Temperature dependence of localization.  Data are shown for the marginally localized case at the lowest temperature with $\Delta$ fixed to $\Delta_c=1$~$E_R$ at $s=4$~$E_R$ (closed circles) and for $\Delta=0.4$~$E_R$ (open circles).  Representative images are shown in false color for gases without an applied impulse for $T_{lat}\approx30$~nK (i) and $T_{lat}\approx70$~nK (ii) for $\Delta=1$~$E_R$. Slices are shown through these images using the scheme from Fig. 2b for $\Delta=1$~$E_R$ (black line) and $\Delta=0.4$~$E_R$ (red line) in (iii).  The gray band corresponds to our resolution limit $v_{res}$ for center-of-mass motion, and the red band is the 95\% confidence interval for a linear fit (red line) to the $\Delta=1$~$E_R$ data. The error bars represent the standard error for the 12--48 experimental runs that were averaged for each data point.}
\label{fig:Fig4}
\end{figure}

In both cases, the motion of the gas is insensitive to temperature.  To quantitatively characterize the temperature dependence in the marginally localized case, we fit the data at $\Delta=1$~$E_R$ to a line.  The 95\% confidence interval for this fit overlaps with $v_{res}$ over the full range of temperatures we sample, implying that the gas remains localized as the temperature is raised.  For these data, $v_{res}$ is at least three orders of magnitude smaller than the temperature and chemical potential of the gas. Furthermore, the chemical potential and temperature at the highest $T_{lat}$ are approximately $2.7t$ and $2t$, leading to minimal occupation at the Brilloun-zone boundary and correspondingly minor occupation of single-particle localized states \cite{McKay2009}.   In this regime, the gas is metallic in the absence of disorder.  The data shown in Fig.~4 are thus an significant constraint that imply many-particle excited states are localized by disorder.

This behavior is consistent with MBL, which predicts that the many-particle eigenstates are localized by disorder across a range of energies in the weakly interacting limit of the Hubbard model.  MBL also predicts that the conductivity, which is analogous to $v_{COM}$ in our system, vanishes across a span of temperature \cite{Basko2006,PhysRevB.75.155111}.  Our measurement is consistent with this absence of thermally activated conductivity---the slope of the linear fit in Fig. 4 is inconsistent with a rise in $v_{COM}$ at the 95\% confidence level. Given the lack of a quantitative prediction in the strongly correlated regime, we cannot rule out non-zero thermal conductivity at a level several times smaller than $v_{res}$.


Future studies in this system may focus on measuring other MBL predictions such as area-laws for entanglement entropy \cite{nayak2013} and the possibility that a many-body localized state may fail to thermalize, a situation that has profound consequences for the eigenstate thermalization hypothesis \cite{PhysRevA.43.2046,PhysRevE.50.888}.  Transport phenomena in the MI phase present at higher $s$ can also be measured.  SDMFT for this system predicts that disorder transforms the MI phase into a disordered correlated metal \cite{Semmler2010}.  Probing such changes in transport properties using the impulse method is complicated in the MI regime by the coexisting metallic shell and localized single-particle states \cite{McKay2009} that are occupied when $k_BT_F>12t$.  Alternatively, methods that use a chemical potential imbalance imposed across a channel may be employed in the future to explore transport in the MI regime \cite{Brantut2012} and to probe if disorder leads to non-Fermi-liquid behavior in the metallic regime \cite{Miranda2005}.

\begin{acknowledgements}
We thank Vito Scarola for assistance with the Hartree-Fock and high-temperature series expansion calculations.  We acknowledge funding from the DARPA OLE program, Army Research Office, and the National Science Foundation.
\end{acknowledgements}

\bibliography{LatticeBib}

\end{document}